\documentclass[10pt, conference, compsoc]{IEEEtran}

\IEEEoverridecommandlockouts

\usepackage{amsmath,amssymb,amsfonts} 
\usepackage{graphicx}
\usepackage{textcomp}
\usepackage{xcolor}
\usepackage{enumitem}
\usepackage{algorithm} 
\usepackage{algpseudocode}
\usepackage{booktabs} 
\usepackage{mathrsfs}
\usepackage{subcaption}
\usepackage{braket} 
\usepackage{caption} 
\usepackage{listings}
\usepackage{tabularx} 

\usepackage[version=4]{mhchem} 
\usepackage[T1]{fontenc}

\def\BibTeX{{\rm B\kern-.05em{\sc i\kern-.025em b}\kern-.08em
    T\kern-.1667em\lower.7ex\hbox{E}\kern-.125emX}}

\usepackage{xspace}
\newcommand{\ours}{QChem-Trainer\xspace}
\newcommand{\myeqref}[1]{\eqref{#1}}

\begin{document}

\title{Large-scale Neural Network Quantum States for \textit{ab initio} Quantum Chemistry Simulations on Fugaku\\
}

 

\author{\IEEEauthorblockN{1\textsuperscript{st} Hongtao Xu}
\IEEEauthorblockA{\textit{School of Advanced Interdisciplinary Sciences}\\
\textit{University of Chinese Academy of Sciences}\\
\textit{State Key Lab of Processors}\\
\textit{Institute of Computing Technology, CAS}\\
Beijing, China \\
xuhongtao22@ict.ac.cn}
\and
\IEEEauthorblockN{2\textsuperscript{nd} Zibo Wu}
\IEEEauthorblockA{
\textit{College of Chemistry}\\
\textit{Beijing Normal University}\\
Beijing, China \\
202231150051@mail.bnu.edu.cn}
\and
\IEEEauthorblockN{3\textsuperscript{th} Mingzhen Li}
\IEEEauthorblockA{\textit{State Key Lab of Processors} \\
\textit{Institute of Computing Technology, CAS}\\
Beijing, China \\
limingzhen@ict.ac.cn}
\and
\IEEEauthorblockN{4\textsuperscript{th} Weile Jia}
\IEEEauthorblockA{\textit{State Key Lab of Processors} \\
\textit{Institute of Computing Technology, CAS}\\
Beijing, China \\
jiaweile@ict.ac.cn}
}

\maketitle

\begin{abstract}
Solving quantum many-body problems is one of the fundamental challenges in quantum chemistry. While neural network quantum states (NQS) have emerged as a promising computational tool, its training process incurs exponentially growing computational demands, becoming prohibitively expensive for large-scale molecular systems and creating fundamental scalability barriers for real-world applications. To address above challenges, we present \ours, a high-performance NQS training framework for \textit{ab initio} electronic structure calculations. First, we propose a scalable sampling parallelism strategy with multi-layers workload division and hybrid sampling scheme, which break the scalability barriers for large-scale NQS training. Then, we introduce multi-level parallelism local energy parallelism, enabling more efficient local energy computation. Last, we employ cache-centric optimization for transformer-based \textit{ansatz} and incorporate it with sampling parallelism strategy, which further speedup up the NQS training and achieve stable memory footprint at scale. Experiments demonstrate that \ours accelerate NQS training with up to 8.41x speedup and attains a parallel efficiency up to 95.8\% when scaling to 1,536 nodes.
\end{abstract}


\IEEEpeerreviewmaketitle

\section{Introduction} 
Efficiently solving the electronic Schrödinger equation has long been a key challenge for quantum many-body physics and quantum chemistry. Various standard methods have been proposed for solving the electronic Schrödinger equation such as full configuration interaction (FCI) \cite{sherrill1999configuration}, coupled cluster (CC) \cite{bartlett2007coupled}, and second-order M\o ller–Plesset perturbation (MP2) theory \cite{cremer2011mollerMP2}. However, due to the exponential growth of computational complexity with the electronic degrees of freedom, only a limited number of systems can be solved exactly. 
In addition to these deterministic methods, variational Monte Carlo (VMC) approach \cite{sorella2005wave}, which optimizing trial wavefunctions through stochastic sampling, provides an effective alternative. Recently, the neural network quantum states (NQS) \cite{Carleo2017, Hermann2023_reviews} have emerged as a powerful tool to handle the quantum many-body problem. NQS try to overcome the curse of dimensionality inherent in many-body quantum systems by leveraging the expressive power of neural networks to parametrize quantum state with neural network. In 2017, Carleo and Troyer \cite{Carleo2017} first utilize restricted Boltzmann machines (RBMs) to describe prototypical interacting spin models \cite{Carleo2017}. Various deep learning architectures, including recurrent neural networks (RNNs) \cite{hibat-allah_recurrent_2020, wu2023tensor},convolutional neural networks (CNNs) \cite{choo2019twoCNN,wang2024variational} and transformers \cite{wu_nnqs_sc23, viteritti2023transformer}, have been integrated into NQS to further extending the accuracy to quantum many-body systems. Among them, transformer-based architecture exhibits huge potentials in capture complex correlations in quantumn chemistry systems \cite{selfattn_nqs_2022_ICLR,wu_nnqs_sc23}. 

Despite rapid developments, NQS training still faces several challenges especially when applied to large-scale molecular systems. First, the sampling process of NQS training involves enormous samples, leading to substantial computations and memory footprints. Additionally, the auto-regressive property of sampling incurs dynamically increased amount of samples, resulting significant scalability challenges when extending NQS to modern HPC systems. Worse still, due to the exponential growth of complexity of Hamiltonian, the local energy calculation exhibits prohibitively expensive computation cost, hendering the simulation of large-scale molecular system. Last but not least, the complexity of transformers also grows quadratically and incurs enormous memory requirement such as Key/Value cache, which is directly proportional to the electron orbitals and number of samples in NQS training.

To address the above challenges, we propose \ours, an innovative high-performance NQS training framework for \textit{ab initio} electronic structure calculations and adapt it to the supercomputer Fugaku \cite{noauthor_supercomputer_nodate}. 
The major innovations of this work can be summarized as follows.
\begin{itemize}[label=\textbullet]
    \item A scalable and efficient sampling parallelism for NQS training, which incorporates multi-stage workload partitioning, density-aware dynamic load balance strategy and memory-stable hybrid sampling scheme. Our parallelism overcomes the scalability barrier and efficiently handles the dynamic memory consumptions inherent in NQS algorithm. 
    \item A multi-level energy calculation parallelism. We fully exploits the parallelism inherent in Hamiltonians through a hybrid parallelization, benefiting not only the NQS training but also other simulations in classical \textit{ab initio} quantum chemistry.
    \item Cache-centric optimization for transformers based \textit{ansatz}. To handle the linearly increased cache size with the number of samples, we propose a dynamic cache management strategy which employs fixed-size cache pooling, selective recomputation and movement optimizations. We incorporate it into our sampling parallelism and achieves stable memory consumptions.  
    \item Case studies of real chemical systems demonstrate that our framework's effectiveness and the capability to scale the NQS on the top supercomputers such as Fugaku.
\end{itemize}

Although our implementation is based on Fugaku, our optimization is architecture independent except the vectorization of Hamiltonians calculation.  We believe \ours can be easily transplanted to other systems, such as GPU systems. Our work opens the door for unprecedentedly large-scale NQS for \textit{Ab initio} quantum chemistry 
.

\section{Backgroud}
\subsection{Variational Monte Carlo (VMC) Algorithm 
}
\label{sec:VMC-eloc}
Given the variational wavefunction \textit{ansatz} $|\Psi_{\theta}\rangle$, the ground state energy $E_{\theta}$ can be written as:
\begin{equation} 
\label{eq:ground_state_energy}
\begin{aligned}
    \langle E\rangle 
    = \frac{\braket{\Psi_{\theta}|\hat{H}|\Psi_{\theta}}}{\braket{\Psi_{\theta}|\Psi_{\theta}}} 
    = \frac{\sum_n |\braket{\Psi_{\theta}|n}|^2 {\frac{\braket{n|\hat{H}|\Psi_{\theta}}}{\braket{n|\Psi_{\theta}}}}}{\sum_n |\braket{\Psi_{\theta}|n}|^2} 
    = \mathbb{E}_{P_\theta}[E_{\rm{loc}}(n)]
    \end{aligned}  
\end{equation}
Here the $\theta$ represents the parameters need to be optimized through VMC algorithm, which is equivalent to nerual network optimization problem in NQS. $\hat{H}$ represents the second quantized Hamiltonian in quantum chemistry, which can be written as:
\begin{equation} 
\label{eq:second_quantized_Hamiltonian}
    \hat{H} = \sum_{pq}h_{pq}\hat{a}_p^{\dagger}\hat{a}_q + \frac{1}{4} \sum_{pqrs}\braket{pq\| rs}\hat{a}_{p}^{\dagger}\hat{a}_q^{\dagger} \hat{a}_s\hat{a}_r 
\end{equation}
The matrix element $H_{nm}$ in $\hat{H}$ is defined as $\braket{n|\hat{H}|m}$ where the $\ket{n}$ is the occupation number vector (ONV). 
Specifically, the ONV $\ket{n}$ is expressed as $\ket{n_{1\alpha},n_{1\beta},\dots,n_{K\alpha},n_{K\beta}}$ with $n_{k\sigma}\in \{0, 1\}$, where $n_{k\alpha}$ and $n_{k\beta}$ represent the occupation number of the \textit{alpha} and \textit{beta} spin orbital with the same spatial orbital, respectively. The probability distribution ${P_{\theta}(n)}$, which can be obtained using Markov Chain Monte Carlo (MCMC) Sampling, can be expressed as:
\begin{equation} 
    P_{\theta}(n) = \frac{|\braket{\Psi_{\theta}|n}|^2}{\sum_n |\braket{\Psi_{\theta}|n}|^2} 
\end{equation}
As shown in \myeqref{eq:ground_state_energy}, the ground state energy can be evaluated using the average of the local energy.
Furthermore, we could evaluate the gradient of the energy as follows:
\begin{equation} 
\label{eq:energy_gradient}
    \partial_{\theta}\langle E\rangle = 2\Re\big[ \big\langle(\partial_\theta\ln{\Psi^*(n)}\rangle) (E_{\rm {loc}}-\langle E\rangle) \big\rangle_n \big] 
\end{equation}
where $\partial_\theta\ln{\Psi^*(n)}$ is the gradient of the parameters $\theta$, which can be computed using the standard automatic differentiation techniques.
The parameters $\theta$ are updated based on the $\partial_{\theta}\langle E\rangle$ with appropriate optimizer, such as SGD, Adam, AdamW and K-FAC \cite{robbins1951stochastic, kingma2014adam, Loshchilov2019, martens2015optimizing}.

\subsection{NQS Workflows 
}
\label{sec:NNQS}
\begin{figure}[htbp]
\centering
\hspace*{-0.3cm}
\includegraphics[width=0.50\textwidth]{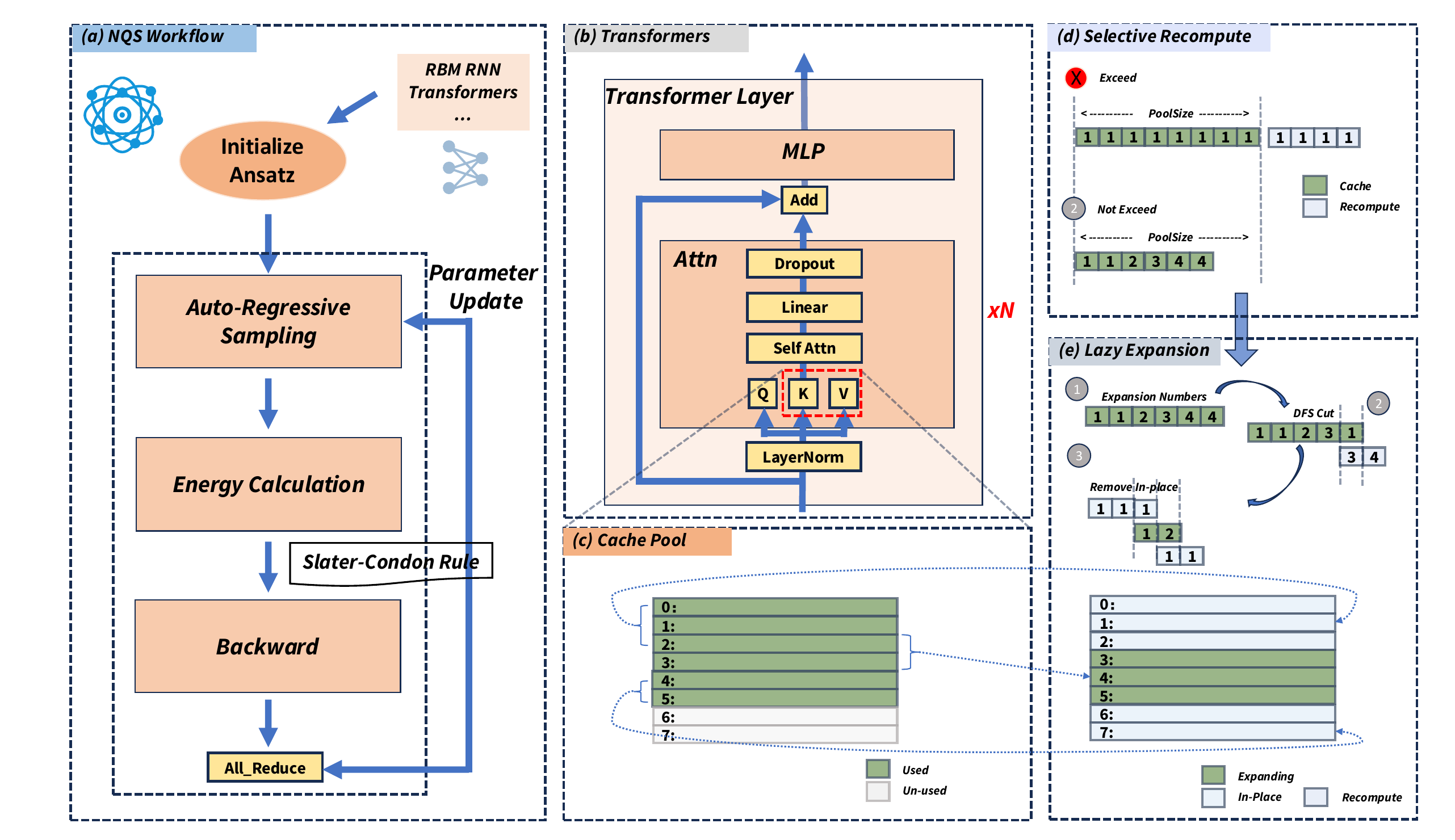}
\caption{(a): NQS workflows. (b): Transformers architecture. (c-e): Cache-centric optimization.} \label{fig:overview-kvcache} 
\end{figure}

As shown in Fig.~\ref{fig:overview-kvcache}a, the NQS workflows comprise three components: auto-regressive sampling, energy calculation and backward. In the beginning, NQS initialize the wave function \textit{ansatz} with a customized neural network, such as RBM, RNNs or Transformers. Then, NQS optimize the neural network iteratively. First, NQS employs sampling phase to get the energy estimation at current state, which mirrors the inference of neural network. Next, NQS compute the energy estimations of the obtained samples, which involves substantial calculation of Hamiltonians. Last, as shown in \myeqref{eq:energy_gradient}, NQS evaluate the approximate gradient by minimize the ground state energy, followed by optimizer step to update parameters. 

In sampling phase, NQS mirrors the network inference except for the total probability property. 
Unlike traditional language models decoding, which predict a probability distribution and select the most probable token as the next output, NQS sampling phase take all the probability into account. Specifically, NQS conducts stochastic sampling with a fixed number of samples ($N_{count}$) under a probability distribution given by one step neural network inference. This approach theoretically leads to an increase in sample size by a factor of $O(V^{N})$, where $N$ is sample steps and $V$ denotes the probability table size, typically represents four possible 
occupation patterns corresponding respectively to the states $\{ \ket{\rm{vac}}, \ket{\alpha}, \ket{\beta}, \ket{\alpha\beta}\}$. Furthermore, chemical-informed prouning is employed to remove the invalid states \cite{zhao_scalable_2023} during every sampling steps.



\section{Innovations} 
\subsection{Scalable and Memory-stable Sampling Parallelism}

In practice, the number of unique sample ($N_u$) represents the actually workloads during each sampling step (similar to the batch size of nerual network inference). However, as mentioned in Section~\ref{sec:NNQS}, NQS sampling phase faces exponential increased samples, leading to vast $N_u$ and huge memory requirements. Worse still, due to the parameter evolution and the randomness of inherent in NQS sampling, $N_u$ is dynamic changed during each iteration. Those characteristics incur huge and unpredictable memory requirements, posing significant barriers for large-scale NQS training.


To address those challenges, we propose a scalable and memory-stable sampling parallelism, which comprises three core components: (i) Multi-stage workload partitioning, which overcomes the barrier of scaling to massively parallel systems. (ii) Density-aware load balance, which tackles the load imbalance problem caused by the dynamic change of $N_u$ through historical information. (iii) Hybrid sampling strategy, which combines breadth-first search (BFS) and depth-first search (DFS) sampling schemes to constrain the peak memory consumptions and enhance the stability during large scale training. 
\subsubsection{Multi-stage Workload Partitioning}
\label{sec:multi-split}
\begin{figure}[htbp]
\centering
\vspace{-5mm}
\includegraphics[width=0.5\textwidth]{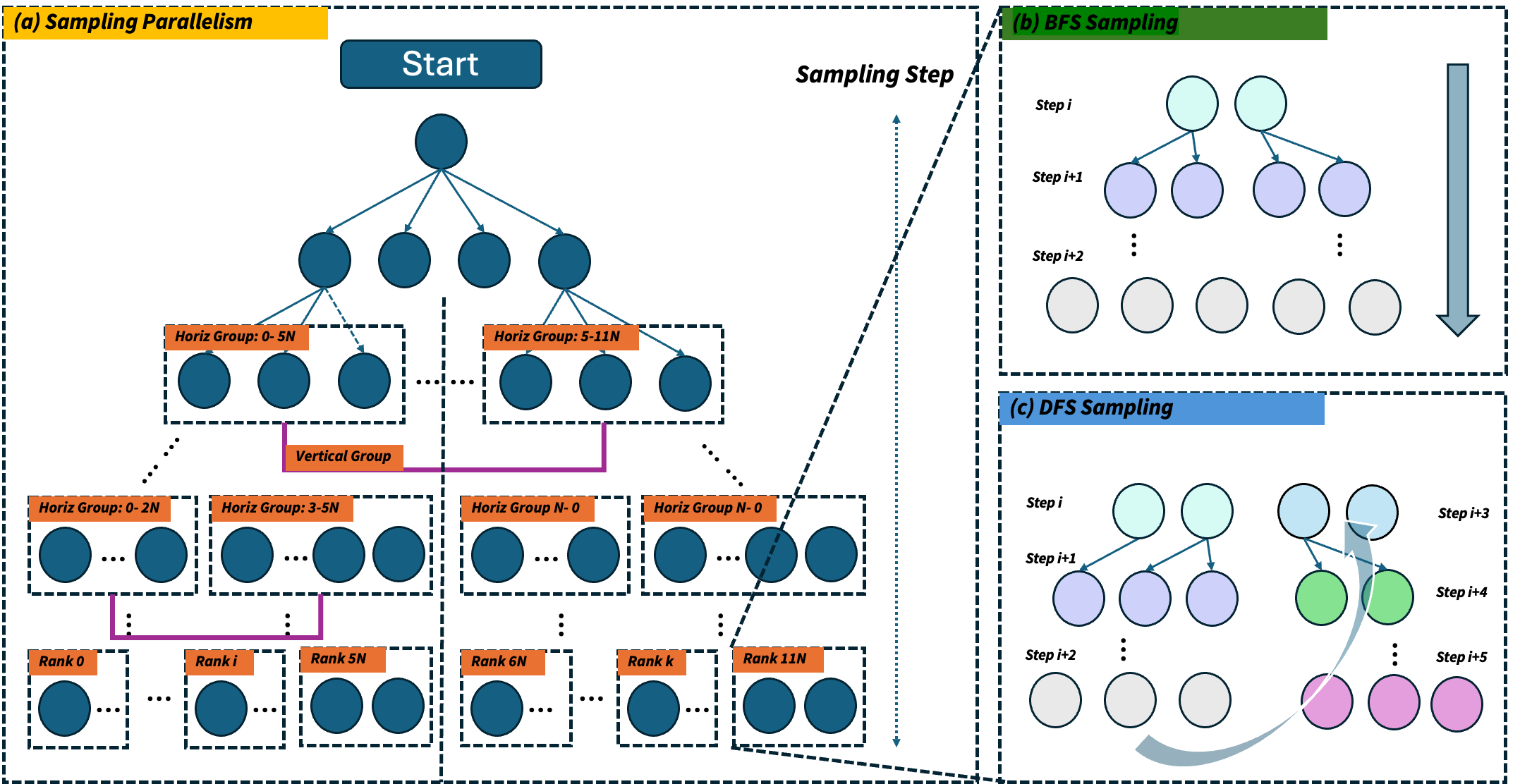}
\caption{Overview of the sampling parallelsim. (a): Multi-stage workload paritioning. (b): Default BFS sampling scheme. (c): DFS sampling scheme.} \label{fig:sampling}
\end{figure}
As shown in Fig.~\ref{fig:sampling}a 
, the NQS sampling phase can be regarded as a quadtree, where the tree layers and nodes represent auto-regression sampling steps and unique samples, respectively. We can simply implement an embarrassingly parallel approach where $N_p$ processes samples independently. However, this approach introduces significant redundant samples, resulting in low end-to-end training efficiency. A straightforward solution to the redundancy issue is to fix the random seed, ensuring that each process generates an identical quadtree. Subsequently, the tree nodes at a specific $k-th$ layer are parallelized \cite{wu_nnqs_sc23}. However, this method still faces a scalability barrier in large-scale parallelization. The root cause lies in its inherent requirement for all processes to sample the entire set of unique samples ($N_u$) at the $k-th$ layer before distributing the workloads. $N_u$ must satisfy $N_u \geq cN_p$ to achieve load balance across the processors, where $c$ is a constant. As $N_p$ increases, $N_u$ grows proportionally, quickly exceeding the memory capacity of a single device. This limitation makes this one-stage partitioning method unable to scale effectively on modern supercomputers.

\indent To address this problem, we propose a multi-stage workload partitioning strategy for NQS, as illustrated in Fig.~\ref{fig:sampling}a 
. Contrast to one-time workload division, our method organizes processes into multiple groups and gradually partitions the workload within each group. This design effectively mitigates the challenges posed by the large $N_u$ requirement. Specifically, given a list of process group sizes $G_n$ and a list of split layers $L$, we divide the current workload into $G_n[i]$ parts at layer $l[i]$, where $i$ is the index for $G_n$ and $L$. 
The relationship between the total number of processors ($N_p$) and the partition setting ($G_n$) satisfy $N_p = \prod_{i=1}^{|G_n|} G_n[i]$.


To facilitate this partitioning and density-aware load balance (see in Section~\ref{section:dynamicLB}), we organize the processes into VerticalGroups $V_g$ and HorizGroups $H_g$ at each partition layer. As depicted in Fig.~\ref{fig:sampling}a, VerticalGroups $V_g$ is responds to partition while the HorizGroups is for load balancing, detailed in \S\ref{section:dynamicLB}. When current layer $k$ is arrive on split layers $L[i]$, every ranks apply division across $V_g[i]$ according to weights (detailed in \S\ref{section:dynamicLB}). For example, given the $G_n=[2,2,3]$ (as illustrate in Fig.~\ref{fig:sampling}a, the purple line reference vertical group) and $L=[6,8,10]$. In rank 0, $V_g=[[0,6],[0,3],[0,1,2]]$ and $H_g=[[0,1,2,3,4,5],[0,1,2],[0]]$. This hierarchical grouping enables efficient workload distribution and load balance across massively parallel environment. 

\begin{algorithm}[htbp]
\footnotesize
\caption{Process Group Initialization}
\label{algo:init_goups}
\begin{algorithmic}[1]
\State \textbf{Given} current HorizGroup $H_{go}$, split list $L$
\State \textbf{Return $H_{g}$, $V_{g}$}
\State $H_g, V_g=[], []$
\State $H_{go}=get\_default\_group()$
\For{$i=0,\ldots,len(L)-1$}
    \State $ws=H_{go}.get\_wordsize()$
    \State $rank=H_{go}.get\_rank()$
    \State $tmp \gets \lfloor rank/L[i] \rfloor$
    \State $H_{g}.append(init\_group(range(tmp * L[i], (tmp+1) * L[i])))$
    \State $V_{g}.append(init\_group(rank\%L[i]+L[i]*range(ws/L[i]))$)
    \State $H_{go}=H_g[-1]$
\EndFor
\end{algorithmic}
\end{algorithm}

\subsubsection{Density-aware Dynamic Load Balancing}
\label{section:dynamicLB}
The inherent unpredictability of sampling results complicates accurate workload prediction during intermediate layer partitioning. Traditional static partitioning methods, which rely on the known metrics in current layer like unique sample counts or total samples, often result in significant load imbalance, especially at large scales. We address this challenge through a density-aware dynamic load balancing strategy founded on two key observations: (i) 
Continuity of parameter evolution
Neural network parameters evolve gradually through incremental updates during training, resulting in smooth variations in sample distributions between consecutive iterations. This continuity allows for predictable workload patterns. (ii) 
Spatial locality in quantum states
chemically valid configurations exhibit spatial correlation in the sampling quadtree structure due to molecular orbital interactions.

We leverage these properties through a novel density metric ($d$) representing the \textit{unique-to-sample ratio} ($d=\frac{sample\_unique}{sample\_counts}$). This metric enables workload prediction by multiple sample counts (known in current layer) and $d$. As shown in Algorithm~\ref{algo:Multi-split&load-balance} (Lines 6-12), we refine the original static load balance by incorporating this dynamic metric with minimal overhead from local communication within $H_g$. Our hierarchical group organization (Fig.~\ref{fig:sampling}) enables efficient implementation, as discussed in \S\ref{sec:multi-split}.

\begin{algorithm}[H]
\footnotesize
\caption{Multi-stage Workload Partition with Density-aware Load-balance}
\label{algo:Multi-split&load-balance}
\begin{algorithmic}[1]
    \State \textbf{Given} $L, G_n:List[int]; Rank:int; D=1:int$ 
    \State \textbf{Return} $sample\_unique$, $sample\_counts$
    \State $V_g, H_g\leftarrow InitGroup(Rank, G_n)$  
    \For{$i=0,\ldots,len(L)-1$}
        \State $sample\_unique, sample\_counts \gets Sampling(L[i], L[i+1])$
        \State $w\gets sample\_counts$
        \State $D_{avg} \gets AllReduce(D, H_g[i])$
        \State $D\_lst \gets AllGather(D_{avg}, V_g[i])$
        \State $p\_idx \gets Partition(sampe\_counts, V_g[i])$
        \For{$j=0,\ldots,len(p\_idx)-1$}
            \State $w \gets w[p\_idx[j]:p\_idx[j+1]] * D\_lst[j]$
        \EndFor 
        \State $sample\_unique,sample\_counts \gets Partition(w, V_g[i])$
    \EndFor
    \State $Density\gets\frac{sample\_unique}{sample\_counts}$

\end{algorithmic} 
\end{algorithm}

\subsubsection{Memory-stable Hybrid Sampling Strategy}
\label{sec:hybrid_sampling}
To address the challenge of memory instability caused by the combinatorial explosion of samples ($O(V^N)$), we propose a hybrid BFS-DFS sampling strategy with guaranteed memory constraints. As illustrated in Fig.~\ref{fig:sampling}(b-c), this method integrates breadth-first search (BFS) and depth-first search (DFS) approaches to optimize memory usage and ensure stable performance during large-scale training. 
Our proposed method enables automatic switching between BFS and DFS sampling schemes based on the current sample count $N_u$ relative to the threshold 
. Specifically, when $N_u < k$, sampling proceeds at the granularity of tree layers, similar to traditional BFS. Once $N_u \geq k$, the sampling process switches to a DFS approach with a stride of $k$, effectively managing peak memory consumption by processing smaller batches sequentially. In this mode, a stack is maintained to track the string information of samples exceeding the threshold, discarding intermediate caches to further reduce memory footprint. Upon reaching leaf nodes in the quadtree, final sample results are recorded, and the stack is popped to continue the sampling process until completion.
 
\subsection{Multi-level Energy Calculation Parallelism}
\label{subsec:energy_parallel}
The evaluation of local energies (\myeqref{eq:loc_energy}) constitutes the computational bottleneck in NQS training particularly for molecular systems with complex second-quantized Hamiltonians. The Hamiltonian element contains $\mathcal{O}(N^4)$ terms for $N$ spin orbitals, resulting in quartic scaling of computational cost with molecular system size. 
\begin{equation} 
\label{eq:loc_energy} 
    E_{\rm{loc}}(n)  = \sum_m \bra{n}\hat{H}\ket{m}\frac{\Psi(m)}{\Psi(n)}
\end{equation}  

To address this challenge, we develop a multi-level parallelization approach and reform the local energy calculations as Algorithm~\ref{algo:energy_parallel}. We first decompose the energy calculation as \myeqref{eq:multi-level-energy}. First, we parallelize the each ONV $\ket{n}$ terms in MPI-level, which is the unique samples $N_u$ obtained from NQS sampling phase and corresponds to the outer loop in \myeqref{eq:multi-level-energy} 
. Then, we employs threads-level parallelization to further accelerate the calculation, which is the middle loop in \myeqref{eq:multi-level-energy}. Last, to fully exploit the parallelism, we vectorize the inner loop in SIMD level. 
\begin{equation}
\label{eq:multi-level-energy}
E_{\text{loc}} = \frac{1}{N_p}\sum_{k=1}^{N_p}\left(\sum_{n\in\mathcal{N}_k}\left(\sum_m\bra{n}\hat{H}\ket{m}\frac{\Psi(m)}{\Psi(n)}\right)\right)
\end{equation} 

However, the vectorization of the matrix element computation $\bra{n}\hat{H}\ket{m}$ in inner loop is not trivial. 
There lies three key barriers: (i) The number of orbits in chemical system always not fit into a vector register and the number of orbits always differs across different systems. (ii) Conditional branching inherent in Slater-Condon rules \cite{scemama2013efficientimplementationslatercondonrules}, which is our baseline implementation and require three excitation conditions: identical configurations, single excitations, and double excitations. (iii) The complex bit-wise operations and irregular access patterns for two-electron integrals (h2e) arrays.

We overcome these barriers through three optimizations:
\begin{itemize}
\item \textbf{Qubit-packing}: We compress orbital occupations into 64-qubit chunks so that each chunk can be stored as a double float format. Consequently, through interleaved loading double floats into multiple vector registers and carefully dealing with the interaction of those vector registers, we can fully exploit the parallelism in SIMD level. Algorithm~\ref{algo:energy_parallel} line 6-7 demonstrate one chunk situation.
\item  \textbf{Branch Elimination}: In practical chemical systems, the double excitation state dominates the calculation. Based on this chemistry insight, we remove the conditional branches for more thorough vectorization and efficient execution pipeline and implement customized check function to ensure the correctness (see Algorithm~\ref{algo:energy_parallel} line 12-14). In this case, some common calculation can be fused such as creations (annihilations) state counting and h2e access (see Algorithm~\ref{algo:energy_parallel} line 8-11).
\item \textbf{Other SIMD Patterns}: Implement other frequently used component, including parity calculation (see Algorithm~\ref{algo:energy_parallel} line 9).
\end{itemize}
Experiment shown in \S\ref{sec:evaluation} demonstrate our methods is highly efficient and scalable.

\begin{algorithm}[htbp]
\footnotesize
\caption{Three-level Parallelism in Energy Calculation}
\label{algo:energy_parallel}
\begin{algorithmic}[1]
\State Partition samples across MPI processes: $\mathcal{M} = \bigcup_{k=1}^{N_p}\mathcal{M}_k$
\State Each process:  
\For{$n \in \mathcal{N}_k$ in OpenMP parallel} \Comment{Thread-level Paralelism}
    \State Compute $\bra{n}\hat{H}\ket{m}$ via SVE-vectorized: 
    \For{$i = 0 \text{ to } N_{\text{vec}} \text{ with stride } \texttt{svcnt}$}
        \State $\mathbf{sv\_bra} \gets \textbf{sv\_dup}(n)$ \Comment{ Broadcast bra state}
        \State $\mathbf{sv\_ket} \gets \textbf{svld1}(m[i])$ \Comment{ Load svcnt ket state} 
        \State $\mathbf{sv\_p}, \mathbf{sv\_q}, \mathbf{sv\_n} \gets \textbf{sv\_fused\_bitop}(sv\_bra, sv\_ket)$
        \State $\mathbf{sv\_sgn} \gets \textbf{sv\_parity}(sv\_bra, sv\_ket, sv\_p, sv\_q)$
        \State $\mathbf{sv\_Hni} \gets \textbf{sv\_mul}(\textbf{sv\_sgn}, \textbf{sv\_h2e}(h2e, sv\_p, sv\_q))$
        \State $\textbf{sv\_store}(\textbf{sv\_Hni})$
        \State $\mathbf{pred\_0} \gets \textbf{sv\_cmpeq}(sv\_n, 0)$ \Comment{Identity other case}
        \If{$\textbf{sv\_ptest\_any(pred\_0)}$} 
            \State Customized\_function()
        \EndIf
    \EndFor
\EndFor
\State Reduce local energies: $\text{MPI\_Allreduce}(E_{\text{loc}}^{(k)})$
\end{algorithmic}
\end{algorithm}

\subsection{Cache-Centric Optimization for Transformers}
\label{subsec:kvcache_optim}
Key/Value cache (KVCache), which stores previous Key/Vaules to avoid redundancy computations when decoding, is the common optimization technique 
\cite{NIPS2017_attentionisallyouneed}
to accelerate the inference of transformer architectures. However, directly applying KVCache to NQS sampling is infeasible. 
Specifically, the size of KVCache is in direct proportion to the $N_u$ and its length (the sampling steps). In practical NQS training, $N_u$ is required to be very large (often exceeds $10^6$) for accurate energy estimation, leading to substantial memory requirements. Furthermore, the dynamic change of $N_u$ in each iteration also increases the uncertainty in memory footprint. Worse still, within the perspective of sampling step, $N_u$ is expanded rapidly with roughly $O(N_{step}^4)$ scaling, resulting in frequent data movements. Therefore, the memory issues of KVCache hinders its applicability in large scale molecular systems. 
To address this problem, we propose a dynamic cache management strategy which employs cache pooling, selective recomputation and lazy cache expansion, achieving stable memory footprint and speedup within limited memory capacity.

\subsubsection{Fixed-size Cache Pooling and Selective Recomputation} 
We first organize the KVCache into a fixed memory pre-allocation pool. Cache pooling strategy not only ensure the controllable peak memory consumption but also avoid the overhead of frequent memory allocations and segmentation caused by samples expanding. Additionally, selective recomputation is employed to handle the situation when the unique samples exceed the cache pool capacity. When the KVCache exceed the capacity, we discard the exceeding part of KVCache and employ recomputation strategy to get the previous Key/Value. Although recomputation increases the computation FLOPs, its side effect remains negligible especially when incorporates with hybrid sampling strategy (detailed in Section~\ref{sec:hybrid_sampling}). Specifically, we reuse the sampling chunk size $k$ as cache line size. The recomputation will be employed only in case of switching to DFS sample scheme, which the sampling chunks' KVCache will be discarded except for the first one. Therefore, our strategy only increase once additional computation in sampling scheme switching layer. 
 
\subsubsection{Lazy Cache Expansion}
We optimize the data movement through lazy cache expansion during samples increasement in NQS training. Due to the property of sampling, each sample step the batch size will expand with the fact of $\lambda$ ($\leq 4$), leading memory reorder in the cache pool. We optimize this process by delaying the KVCache expansion and seeking the underline optimization opportunity, which we named it lazy expansion. As shown in Fig.~\ref{fig:overview-kvcache}(c-d), we just keep the expanding indexs after generating the new unique samples. We optimize the unnecessary memory reordering by analysis the indexs: (i) Detecting over-long expansion by sampling chunk size $k$ (ii) Keep in-place of the leading cache block which expanding size is 1 (iii) Directly in-place moving the least cache block which expanding size is non-zero.

\section{Evaluation}
\label{sec:evaluation} 
\subsection{System Architecture and Evaluation Setup}
We implement \ours on Fugaku supercomputer, which comprises 158,976 nodes with a theoretical peak performance of 537 PFLOPS and currently ranks No.6 and No.1 on the TOP500 list \cite{noauthor_supercomputer_nodate} and HPCG benchmark \cite{noauthor_hpcg_nodate}, respectively. Each computing node is equipped with one A64FX SoC, which consists of 52 cores with 4 dedicated for OS and 48 for computation. The A64FX SoC is organized into four Core Memory Group (CMGs), with each CMG connected to 8GBs local HBM2 memory with 256GB/s bandwith. Moreover, the A64FX architecture supports 512-bit SVE operations, enabling each A64FX SoC to reach a theoretical peak performance of 3.38 TFLOPS at 2.2GHz. 

In evaluation, we use transformers as the wavefunction ansatz for equal comparation. The setting is as the follows. For the amplitude part of wavefunction, we use 8 decoder-only layers with number of attention head $n_{head} = 8$ and hidden size $d_{model} = 64$ \cite{radford2019gpt2language}. For phase, we use three layers MLP with sizes $N*512*512*1$, where $N$ is the number of spin orbits. We use AdamW \cite{loshchilov2018adamwdecoupled} as optimizer and the learn rate is $lr = 1e-2$ which schedules as \myeqref{eq:learning_rate_scheduler} with $n_{warmup} = 2000$ for default.
\begin{equation}
\label{eq:learning_rate_scheduler}
    \eta_t = d_{\text{model}}^{-0.5} \times \min\left( (t + 1)^{-0.5},\ t \times n_{\text{warmup}}^{-1.5} \right)
\end{equation}
All the experiments are performed on Fugaku with a NUMA-aware MPI setting. 
\subsection{Precision and Performance Validation}
\begin{figure}[htbp]
    \centering
    \begin{subfigure}[c]{0.48\columnwidth} 
        \raisebox{-0.75em}{\includegraphics[width=\linewidth]{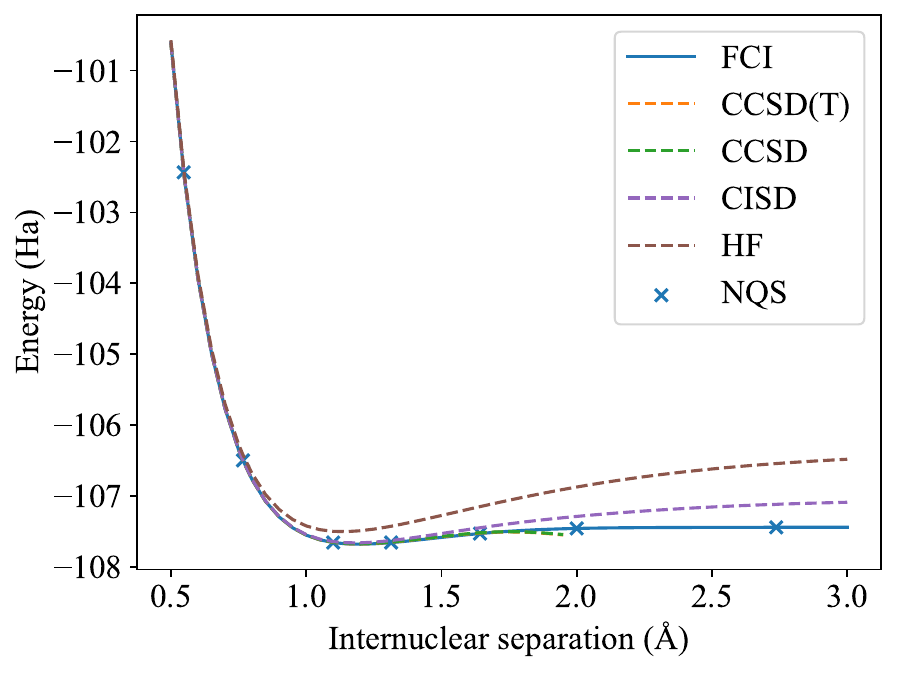}}
    \end{subfigure}
    \begin{subfigure}[c]{0.48\columnwidth} 
        \includegraphics[width=\linewidth]{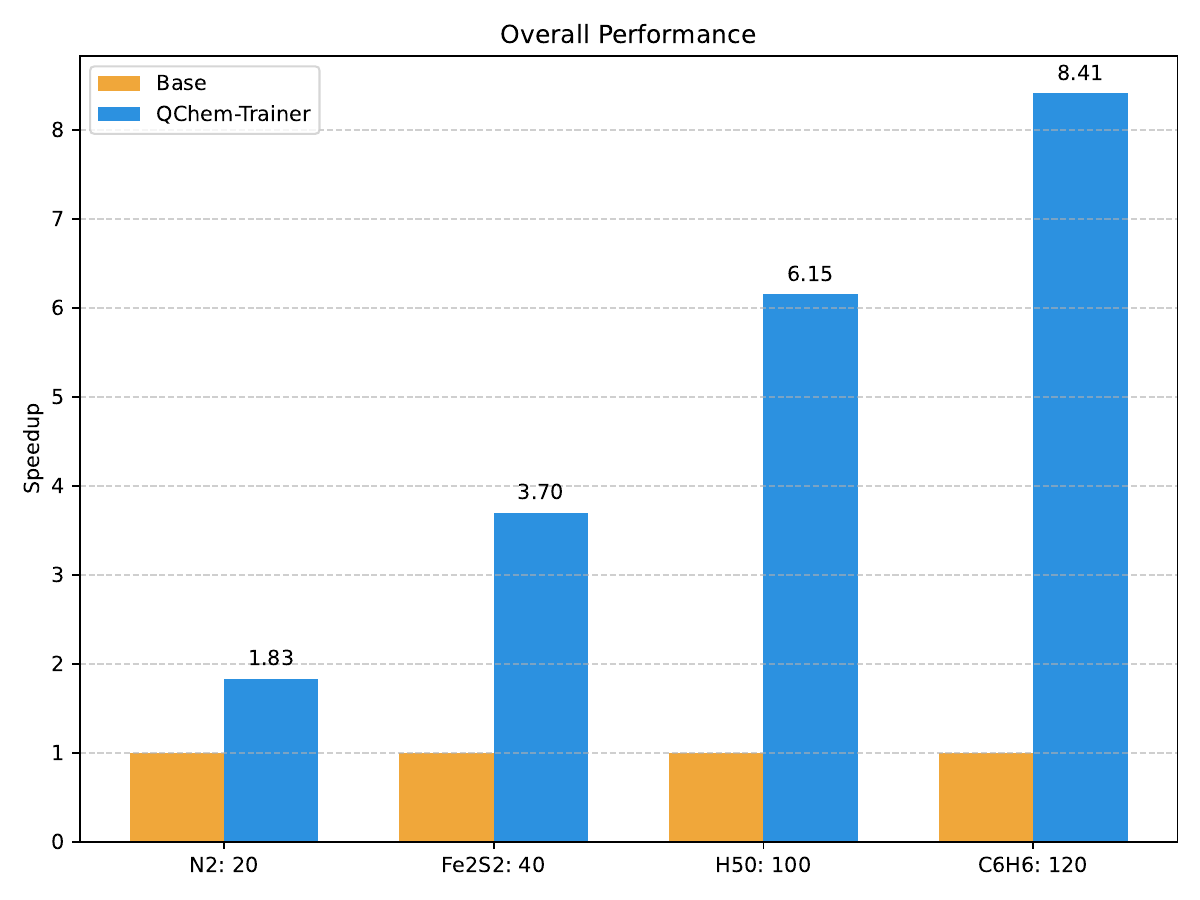}
    \end{subfigure} 
    \caption{Left: Potential energy surfaces of $N_2$. Righ: Overall speedup. Four molecular systems are chosen to test \ours's performance. From left to right, the spin orbits in molecular system increases. The figure demonstrate the normalized speedups compared to baseline implementation.} 
    \label{fig:validation} 
\end{figure}

We evaluate our optimization precision within three real-world molecular systems: \ce{N_2}, \ce{PH_3} and \ce{LiCl} and they are all in STO-3G basis set. As shown in Table~\ref{tab:energies}, \ours can reach the same the ground state energies compared to existing accurate results such as FCI.
Additionally, we calculate the potential energy surface of the $N_2$ molecular system for more comparison, which is shown in Fig.~\ref{fig:validation} Left. Results demonstrate our optimizations precision.

\begin{table}[htbp] 
\centering
\caption{Ground state energies (in Hartree) calculated by our method. The HF, CCSD and FCI results are shown for comparison. $N$ is the number of qubits and $N_e$ the total number of electrons (including spin up and spin down).}
\label{tab:energies}
\begin{tabular}{@{}lrrrcccc@{}}
\toprule
& & & \multicolumn{4}{c}{Energy (Hartree)} \\
\cmidrule(lr){4-7}
Molecule & $N$ & $N_e$ & HF & CCSD & Ours & FCI \\
\midrule
$N_2$     & 20  & 14  & -107.4990 & -107.6560 & -107.6602 & -107.6602 \\ 
$PH_3$    & 24  & 18  & -338.6341 & -338.6981 & -338.6984 & -338.6984 \\
$LiCl$    & 28  & 20  & -460.8273 & -460.8475 & -460.8494 & -460.8496 \\
\bottomrule
\end{tabular}
\end{table}

To test the overall performance gains, we choose four systems with the increased spin orbits: $N_2$, $Fe_2S_2$, $H_{50}$ and $C_6H_6$ with 20, 40, 100, 120 spin orbits respectively. 
Among them, $Fe_2S_2$ is a strongly correlated system \ce{[Fe2S2(SCH3)4]^{2-}} with CAS(30e, 20o), while the hydrogen chain \ce{H50} is configured with a bond length $2.0a_0$ using orthonormalized atomic orbitals obtained in STO-6G basis. $C_6H_6$ is benzene molecular system in the 6-31G basis set.

We illustrate the normalized end-to-end speedup up measured by the execution time per iteration. As shown in Figure~\ref{fig:validation} Right, $C_6H_6$ system attains the maximal speedup (8.41x). $N_2$ system exhibits minimal acceleration with 1.83x speedup. An average 4.95x speedups are achieved over those systems, accelerating the NQS simulation significantly.
Also, we found the systems who have more orbits always exhibit more significant performance gains, which we attribute to the more computational complexity in sampling and Hamiltonian calculations.

\subsection{Case study}
In this section, we conduct some case studies in some real chemical systems. 
We choose \ce{H50}
and \ce{Fe2S2} 
to test our sampling parallelism and density-aware load balance, respectively. For energy parallelism, we use \ce{N2}, \ce{Fe2S2} and \ce{H50} to test step by step speedup with 20, 40 and 100 qubits, respectively. We use two methods of energy calculations, which is sample space calculation and accurate calculation, to test the weak scaling of our methods on 1,536 nodes.


\subsubsection{Sampling Parallelism} 
As shown in Fig.~\ref{fig:memory_stable}, we compare three sampling method: (i) baseline implementation without KVCache optimization; (ii) KVCache version without hybrid sampling scheme and cache managements; and (iii) Memory-stable version which combines hybird sampling and our cache-centric optimization. We incrementally increase the number of samples from an initial $2.5 \times 10^3$ up to $1.024 \times 10^7$ to test peak memory footprints and scalability of different methods. The results demonstrate that the KVCache implementation accelerates the sampling process significantly. However, the memory issue hinders its applicability. We can see the KV cache version occurs OOM error in $2\times10^4$ while the base version in $4\times10^4$. In contrast, our memory-stable methods can constrain the peak memory footprint and scale up to $1.024\times10^7$ samples (three orders of magnitude larger) while maintain efficient performance. 

\subsubsection{Load balancing}
\begin{figure}[htbp]
    \centering
    \begin{subfigure}[b]{0.48\textwidth} 
        \includegraphics[width=\linewidth]{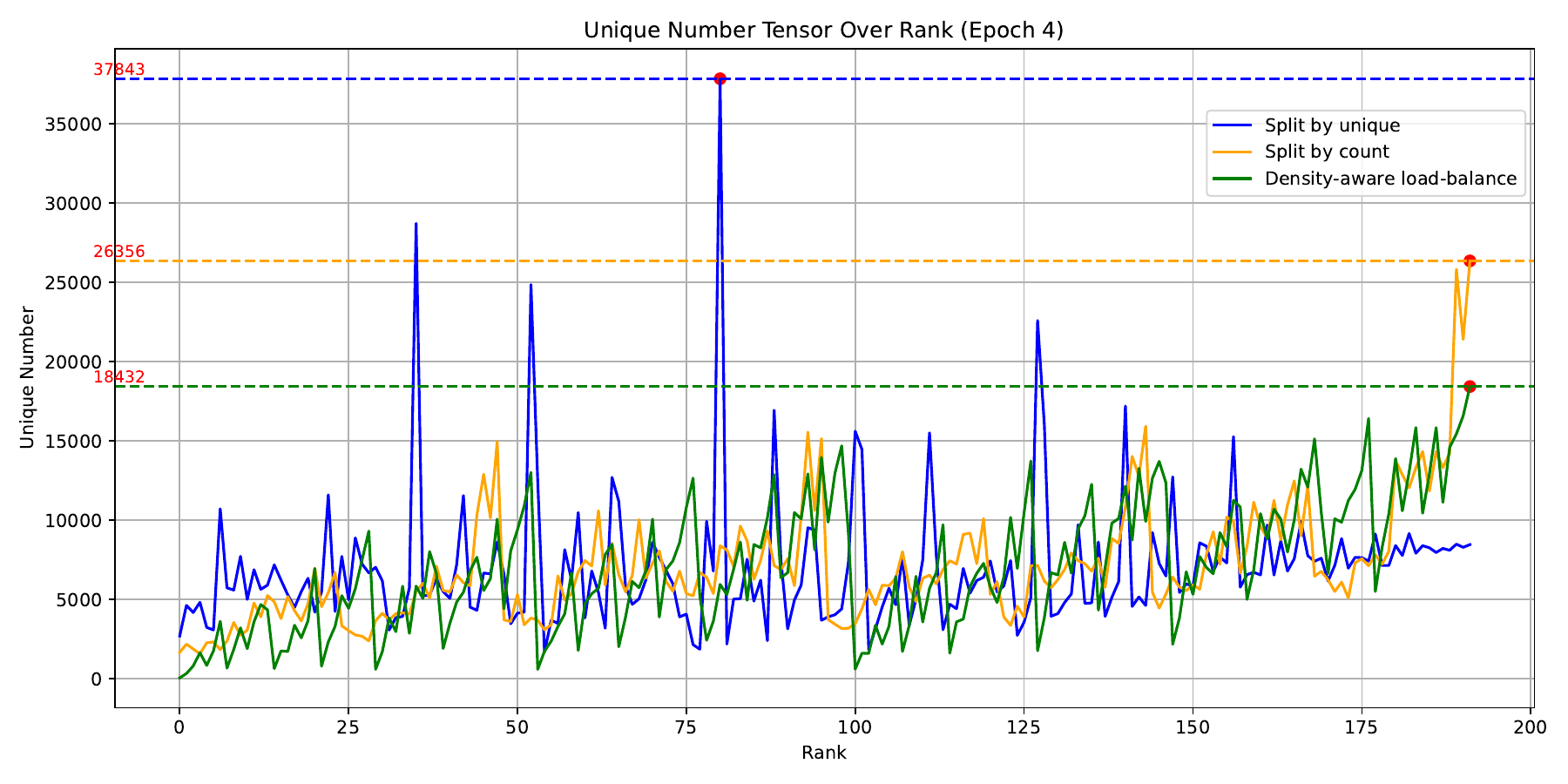}
        \caption{This figure demonstrates the final obtained unique samples $N_u$ in each rank. The blue line denotes splitting by unique samples while the yellow line denotes splitting by sample counts and the green line represents our desity-aware load-balance strategy. 
        }
        \label{fig:load_balance}
    \end{subfigure}
    \hfill 
    \begin{subfigure}[b]{0.48\textwidth} 
        \includegraphics[width=\linewidth]{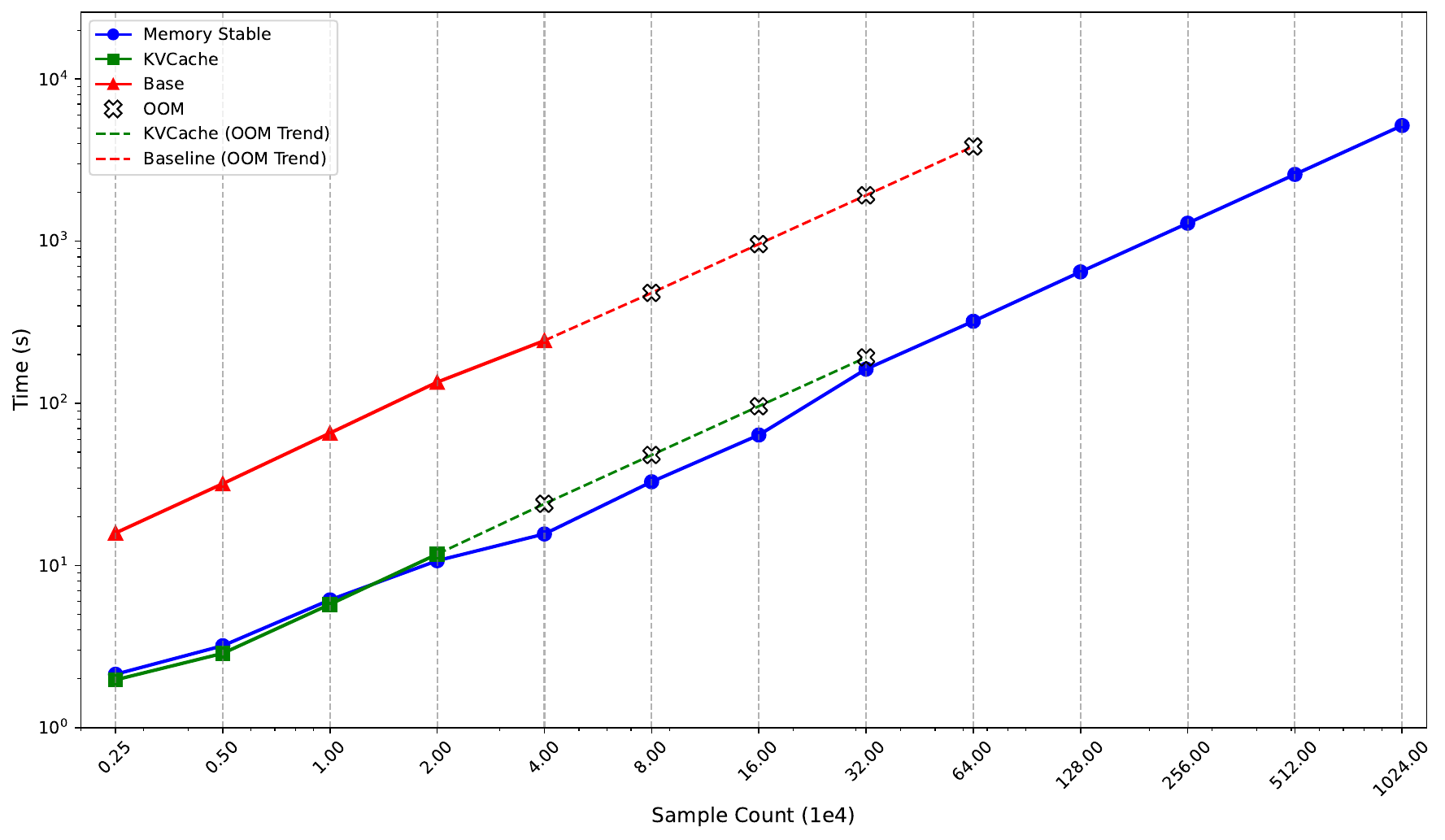}
        \caption{This figure demonstrates the iteration time under the different sample counts. The read line denotes the baseline implementation. The green line denotes directly using KVCache. The blue line represent our memory-stable approach combining both hybrid sampling scheme and cache management. The white cross denotes Out-of-memory (OOM) error.}
        \label{fig:memory_stable}
    \end{subfigure} 
    \label{fig:combined_sampling_parallelism} 
\end{figure}

We collecte the final obtained unique samples to investigate the load balance problem. We use strongly correlated system \ce{Fe2S2} system using 256 ranks and test three methods using the same training settings. Therefore, we can use the maximum unique samples across ranks to demonstrate the real workload situation in NQS training. Experiment results are shown in Fig.~\ref{fig:load_balance}. 
We test two static workload division methods (splited by unique sample and by sample count) and ours density-aware load balancing method, which have 37843, 26356 and 18432 maximum unique samples across all ranks respectively.
As the blue line shown, evenly distributing the unique samples leads to severe load imbalance problem. Additionally, division along sample counts (the yellow line in Fig.~\ref{fig:load_balance}) mitigate the problem. In contrast, our density-aware load balancing method (The green line) showcase the approximately load balance, enabling the large scale NQS training.

\subsubsection{Energy Parallelism} 
\begin{figure}[htbp]
    \centering
    \includegraphics[width=0.5\textwidth]{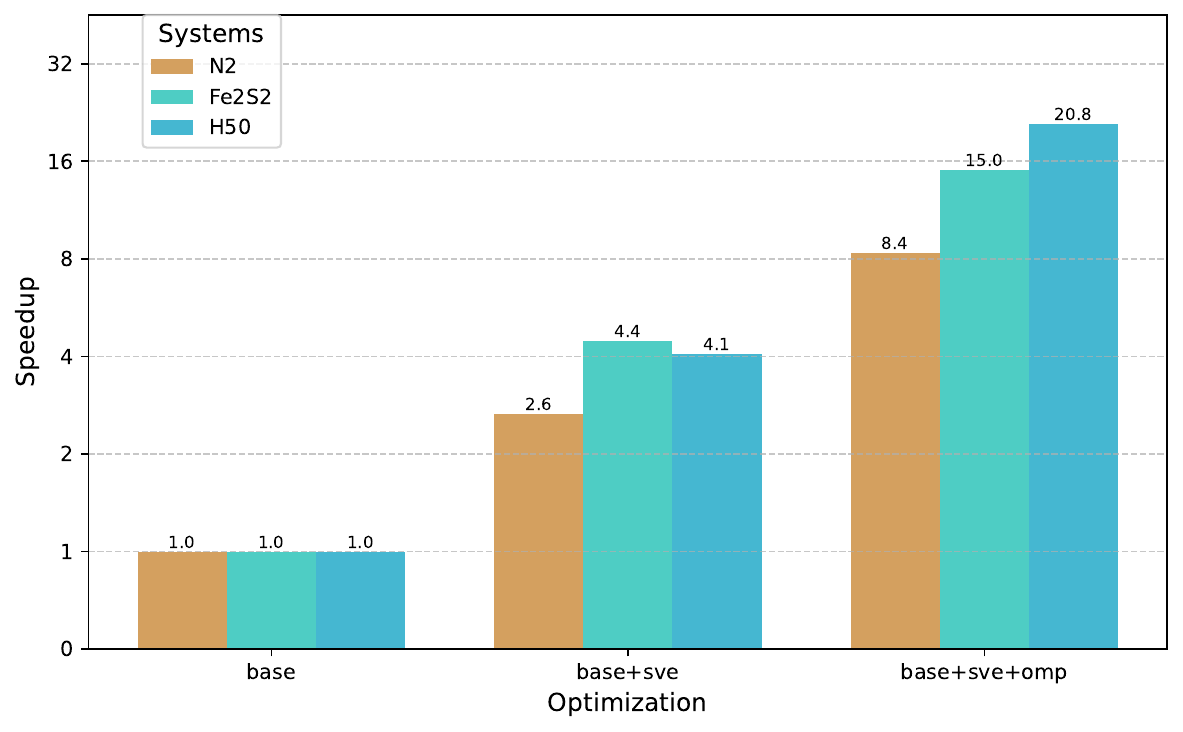}
    \caption{Step by step energy calculation speedup with three chemical systems. Base+sve stands for enabling SIMD optimization while base+sve+omp stands for enabling both of SIMD and thread-level parallel.} 
    \label{fig:energy_speedup} 
\end{figure}

We choose three molecular systems: \ce{N2}, \ce{Fe2s2} and \ce{H50} with increased number of spin orbits to test the step by step speedup of our energy calculation optimization within a single rank. As shown in Fig.~\ref{fig:energy_speedup}, we sequentially enable SIMD (implemented by ARM SVE extension) and thread-level parallel (implemented by OpenMP). 
We can observe the up to 20.8x speedup for \ce{H50} when enable both parallelism (version base+sve+omp). 

\subsubsection{Scalability} 
\begin{figure}[htbp]
    \centering
    \begin{subfigure}[b]{0.48\textwidth} 
        \includegraphics[width=\linewidth]{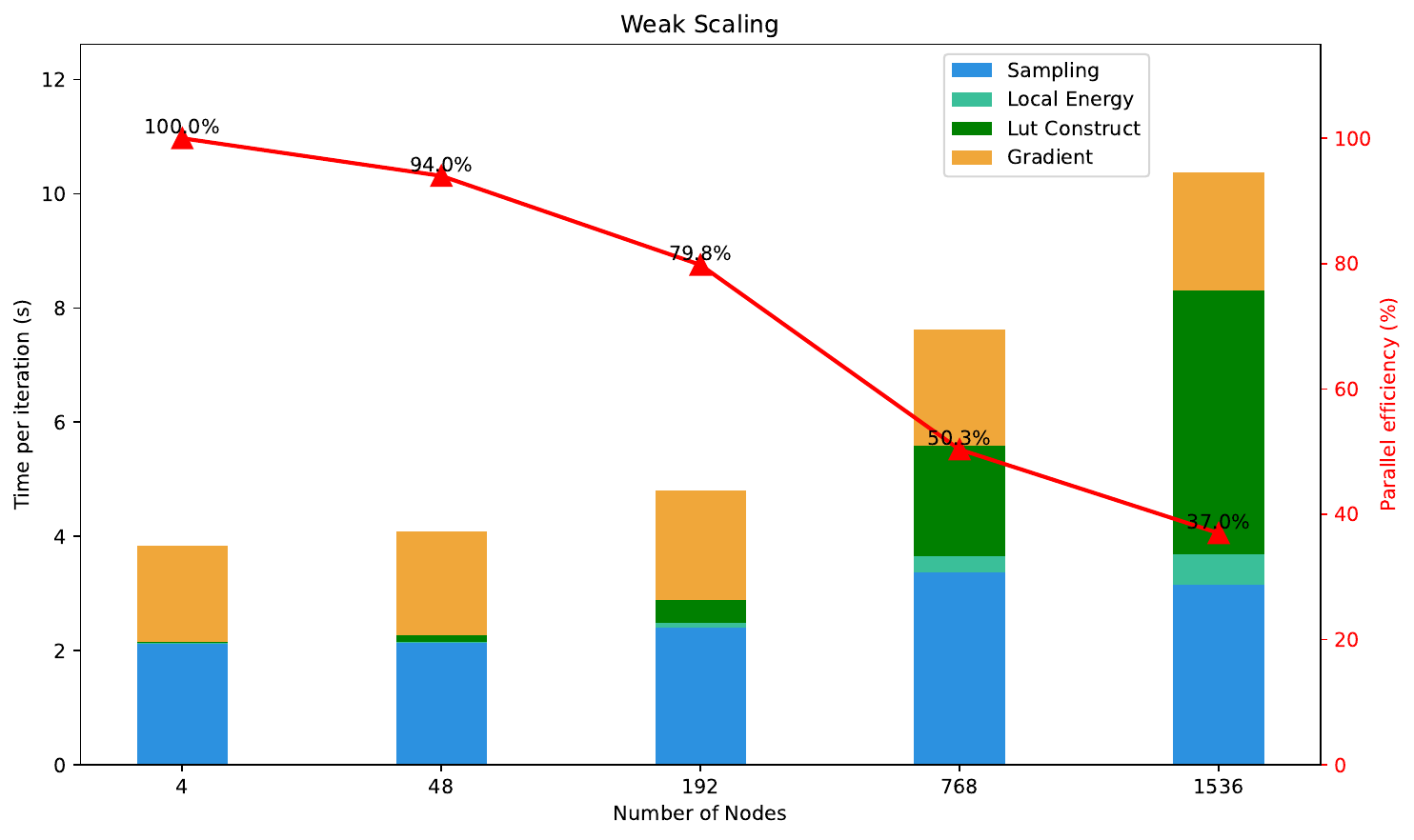}
        \caption{Using sample space energy. }
        \label{fig:weak_scaling_1}
    \end{subfigure}
    \hfill 
    \begin{subfigure}[b]{0.48\textwidth} 
        \includegraphics[width=\linewidth]{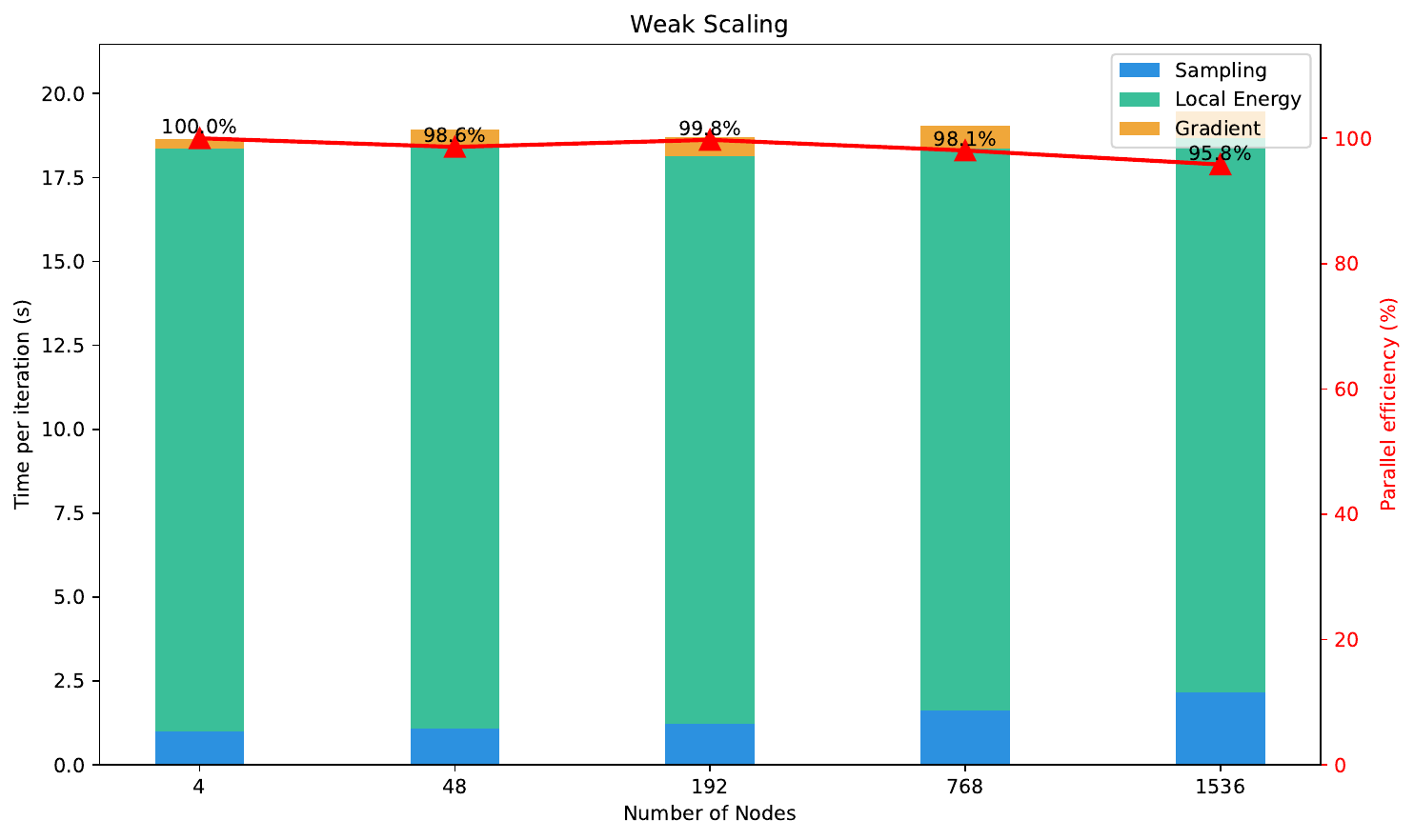}
        \caption{Using accurate energy}
        \label{fig:weak_scaling_2}
    \end{subfigure} 
    \caption{We test the scaling performance in \ce{H50} system with two energy calculation methods. Figure (a) demonstrate the scaling result using sample space energy calculation. In contrast, Figure (b) employs another accurate energy calculation method. Both of them showcase the scalability of \ours.} 
    \label{fig:combined_weak_scaling} 
\end{figure}

We test the scalability of \ce{H50} system with the setting of $N_u = N*4*10^3$ for $N$ nodes. Different $\Psi$ calculation mothods (shown in \myeqref{eq:loc_energy}) are tested for comparision. In sample space methods (shown in Fig.~\ref{fig:weak_scaling_1} ), the overhead of construction of LUT is becoming innegligible when the sample numbers increasing, which is setted to avoid redundant $\Psi$ calculation and noted with green colour, thus degrading scalability. For comparision, we remove the LUT and test the scalability with accurate calculation of $\Psi$, which is shown in Fig.~\ref{fig:weak_scaling_2}. Both results demonstrate our framework scalability in massively parallel NQS training.

\section{Conclusion}
\label{sec:conclusion}
In this work, we presents \ours, a high-performance framework for neural network quantum state (NQS) training, which enables efficient and scalable NQS training on modern HPC platforms. To tackle the scalability and performance bottlenecks in NQS, we propose three key innovations: (i) Scalable and memory-stable sampling parallelism, which overcomes the memory wall and enables NQS training at unprecedented scales; (ii) Multi-level energy calculation parallelism, which fully exploits the parallelism and significantly accelerate the energy evaluation especially for complex molecular systems; and (iii) Cache-centric optimization for transformer-based \textit{ansatz}, which efficiently manages of Key/Value caches to achieve controllable peak memory consumptions. 
Extensive experiments on real-world chemical systems demonstrate that \ours achieves up to 8.41× speedup in overall NQS training. Moreover, strong scaling tests show that \ours maintains parallel efficiency of up to 95.8\% across 1,526 nodes, highlighting its scalability and efficiency.
This work not only paves the way for large-scale \textit{ab initio} quantum chemistry simulations, but also offers practical insights into optimizing such hybrid HPC-AI scientific workloads on next-generation exascale systems.




\bibliography{nqs.bib}

\begin{thebibliography}{10}
\providecommand{\url}[1]{#1}
\csname url@samestyle\endcsname
\providecommand{\newblock}{\relax}
\providecommand{\bibinfo}[2]{#2}
\providecommand{\BIBentrySTDinterwordspacing}{\spaceskip=0pt\relax}
\providecommand{\BIBentryALTinterwordstretchfactor}{4}
\providecommand{\BIBentryALTinterwordspacing}{\spaceskip=\fontdimen2\font plus
\BIBentryALTinterwordstretchfactor\fontdimen3\font minus \fontdimen4\font\relax}
\providecommand{\BIBforeignlanguage}[2]{{%
\expandafter\ifx\csname l@#1\endcsname\relax
\typeout{** WARNING: IEEEtran.bst: No hyphenation pattern has been}%
\typeout{** loaded for the language `#1'. Using the pattern for}%
\typeout{** the default language instead.}%
\else
\language=\csname l@#1\endcsname
\fi
#2}}
\providecommand{\BIBdecl}{\relax}
\BIBdecl

\bibitem{sherrill1999configuration}
C.~D. Sherrill and H.~F. Schaefer~III, ``The configuration interaction method: Advances in highly correlated approaches,'' in \emph{Advances in quantum chemistry}.\hskip 1em plus 0.5em minus 0.4em\relax Elsevier, 1999, vol.~34, pp. 143--269.

\bibitem{bartlett2007coupled}
R.~J. Bartlett and M.~Musia{\l}, ``Coupled-cluster theory in quantum chemistry,'' \emph{Reviews of Modern Physics}, vol.~79, no.~1, pp. 291--352, 2007.

\bibitem{cremer2011mollerMP2}
D.~Cremer, ``M{\o}ller--plesset perturbation theory: from small molecule methods to methods for thousands of atoms,'' \emph{Wiley Interdisciplinary Reviews: Computational Molecular Science}, vol.~1, no.~4, pp. 509--530, 2011.

\bibitem{sorella2005wave}
S.~Sorella, ``Wave function optimization in the variational monte carlo method,'' \emph{Physical Review B—Condensed Matter and Materials Physics}, vol.~71, no.~24, p. 241103, 2005.

\bibitem{Carleo2017}
G.~Carleo and M.~Troyer, ``Solving the quantum many-body problem with artificial neural networks,'' \emph{Science}, vol. 355, no. 6325, pp. 602--606, 2017.

\bibitem{Hermann2023_reviews}
J.~Hermann, J.~Spencer, K.~Choo, A.~Mezzacapo, W.~M.~C. Foulkes, D.~Pfau, G.~Carleo, and F.~Noé, ``Ab initio quantum chemistry with neural-network wavefunctions,'' \emph{Nature Reviews Chemistry}, vol.~7, no.~10, pp. 692--709, Aug. 2023.

\bibitem{hibat-allah_recurrent_2020}
\BIBentryALTinterwordspacing
M.~Hibat-Allah, M.~Ganahl, L.~E. Hayward, R.~G. Melko, and J.~Carrasquilla, ``Recurrent neural network wave functions,'' \emph{Physical Review Research}, vol.~2, no.~2, p. 023358, Jun. 2020. [Online]. Available: \url{https://link.aps.org/doi/10.1103/PhysRevResearch.2.023358}
\BIBentrySTDinterwordspacing

\bibitem{wu2023tensor}
D.~Wu, R.~Rossi, F.~Vicentini, and G.~Carleo, ``From tensor-network quantum states to tensorial recurrent neural networks,'' \emph{Physical Review Research}, vol.~5, no.~3, p. L032001, 2023.

\bibitem{choo2019twoCNN}
K.~Choo, T.~Neupert, and G.~Carleo, ``Two-dimensional frustrated j 1-j 2 model studied with neural network quantum states,'' \emph{Physical Review B}, vol. 100, no.~12, p. 125124, 2019.

\bibitem{wang2024variational}
J.-Q. Wang, H.-Q. Wu, R.-Q. He, and Z.-Y. Lu, ``Variational optimization of the amplitude of neural-network quantum many-body ground states,'' \emph{Physical Review B}, vol. 109, no.~24, p. 245120, 2024.

\bibitem{wu_nnqs_sc23}
Y.~Wu, C.~Guo, Y.~Fan, P.~Zhou, and H.~Shang, ``Nnqs-transformer: an efficient and scalable neural network quantum states approach for ab initio quantum chemistry,'' in \emph{Proceedings of the International Conference for High Performance Computing, Networking, Storage and Analysis}, ser. SC '23.\hskip 1em plus 0.5em minus 0.4em\relax New York, NY, USA: Association for Computing Machinery, 2023.

\bibitem{viteritti2023transformer}
L.~L. Viteritti, R.~Rende, and F.~Becca, ``Transformer variational wave functions for frustrated quantum spin systems,'' \emph{Physical Review Letters}, vol. 130, no.~23, p. 236401, 2023.

\bibitem{selfattn_nqs_2022_ICLR}
I.~von Glehn, J.~S. Spencer, and D.~Pfau, ``{A Self-Attention Ansatz for Ab-initio Quantum Chemistry},'' \emph{11th International Conference on Learning Representations (ICLR)}, 2023.

\bibitem{noauthor_supercomputer_nodate}
\BIBentryALTinterwordspacing
``Supercomputer {Fugaku} - {Supercomputer} {Fugaku}, {A64FX} {48C} 2.{2GHz}, {Tofu} interconnect {D} {\textbar} {TOP500}.'' [Online]. Available: \url{https://www.top500.org/system/179807/}
\BIBentrySTDinterwordspacing

\bibitem{robbins1951stochastic}
H.~Robbins and S.~Monro, ``A stochastic approximation method,'' \emph{The annals of mathematical statistics}, pp. 400--407, 1951.

\bibitem{kingma2014adam}
D.~P. Kingma and J.~Ba, ``Adam: A method for stochastic optimization,'' \emph{arXiv preprint arXiv:1412.6980}, 2014.

\bibitem{Loshchilov2019}
\BIBentryALTinterwordspacing
I.~Loshchilov and F.~Hutter, ``Fixing weight decay regularization in adam,'' in \emph{International Conference on Learning Representations}, 2019. [Online]. Available: \url{https://openreview.net/forum?id=rk6qdGgCZ}
\BIBentrySTDinterwordspacing

\bibitem{martens2015optimizing}
J.~Martens and R.~Grosse, ``Optimizing neural networks with kronecker-factored approximate curvature,'' in \emph{International conference on machine learning}.\hskip 1em plus 0.5em minus 0.4em\relax PMLR, 2015, pp. 2408--2417.

\bibitem{zhao_scalable_2023}
T.~Zhao, J.~Stokes, and S.~Veerapaneni, ``Scalable neural quantum states architecture for quantum chemistry,'' \emph{Machine Learning: Science and Technology}, vol.~4, Jun. 2023.

\bibitem{scemama2013efficientimplementationslatercondonrules}
\BIBentryALTinterwordspacing
A.~Scemama and E.~Giner, ``An efficient implementation of slater-condon rules,'' 2013. [Online]. Available: \url{https://arxiv.org/abs/1311.6244}
\BIBentrySTDinterwordspacing

\bibitem{NIPS2017_attentionisallyouneed}
A.~Vaswani, N.~Shazeer, N.~Parmar, J.~Uszkoreit, L.~Jones, A.~N. Gomez, L.~Kaiser, and I.~Polosukhin, ``Attention is all you need,'' in \emph{Proceedings of the 31st International Conference on Neural Information Processing Systems}, ser. NIPS'17.\hskip 1em plus 0.5em minus 0.4em\relax Red Hook, NY, USA: Curran Associates Inc., 2017, p. 6000–6010.

\bibitem{noauthor_hpcg_nodate}
\BIBentryALTinterwordspacing
``{HPCG} - {November} 2024 {\textbar} {TOP500}.'' [Online]. Available: \url{https://top500.org/lists/hpcg/2024/11/}
\BIBentrySTDinterwordspacing

\bibitem{radford2019gpt2language}
A.~Radford, J.~Wu, R.~Child, D.~Luan, D.~Amodei, I.~Sutskever \emph{et~al.}, ``Language models are unsupervised multitask learners,'' \emph{OpenAI blog}, vol.~1, no.~8, p.~9, 2019.

\bibitem{loshchilov2018adamwdecoupled}
I.~Loshchilov and F.~Hutter, ``Decoupled weight decay regularization,'' in \emph{International Conference on Learning Representations}, 2019.

\end{thebibliography}


\end{document}